\documentclass[twocolumn,prb]{revtex4}

\usepackage{amsfonts}
\usepackage[T1]{fontenc}
\usepackage{amsmath,amsbsy,amssymb,graphicx}
\usepackage{times}
\let\mathbf=\boldsymbol

\begin{document}

\title{{\Large Majorana Fermions and Multiple Topological Phase Transition}\\{\Large in Kitaev Ladder Topological Superconductors}}
\author{Ryohei Wakatsuki$^1$, Motohiko Ezawa$^1$ and Naoto Nagaosa$^{1,2}$}
\affiliation{$^1$Department of Applied Physics, University of Tokyo, Hongo 7-3-1, 113-8656, Japan}
\affiliation{$^2$Center for Emergent Matter Science (CEMS), ASI, RIKEN, Wako 351-0198,
Japan}

\begin{abstract}
Motivated by the InSb nanowire superconductor system, we investigate a system where one-dimensional topological superconductors are placed in parallel.
It would be simulated well by a ladder of the Kitaev chains.
The system undergoes multiple topological phase transitions, where the number of Majorana fermions changes as a function of the interchain superconducting pairings.
We analytically determine the topological phase diagram by explicitly calculating the topological number and the band structure.
They show even-odd effects with respect to the number of legs of the ladder.
When the relative phase between the inter- and intrachain superconducting pairings is 0 or $\pi$,
the system belongs to the class BDI characterized by the $\mathbb{Z}$ index,
and otherwise it belongs to the class D characterized by the $\mathbb{Z}_2$ index.
This topological class change would be caused by applying the Josephson current or an external magnetic field, and could be observed by measuring the zero-bias differential conductance.
\end{abstract}

\maketitle

\section{introduction}

The Majorana fermion is one of the hottest topics in condensed matter physics\cite{Alicea,Beenakker1,Tanaka}.
Majorana fermions are particles which are their own antiparticles.
Because of their nonlocality and non-Abelian statistics, they are considered to realize exotic phenomena and there are possibilities to encode fault tolerant topological quantum computations.
There are several suggestions for physical systems that support Majorana zero-energy states (MZES)\cite{Fu,Yuval,Lutchyn,Sau,Alicea2,Sato1,Sato2}.
The Kitaev model is the simplest model that realizes the MZES\cite{Kitaev}.
It well describes a one-dimensional nanowire with strong Rashba spin-orbit interaction\cite{Yuval,Lutchyn}.
There are generalizations of the Kitaev model by several authors\cite{Potter,Niu,Rieder,Kells,Asahi,Zhou,Manmana,DeGottardi,Wimmer,Gangadharaiah}.

Thin films and multichannel nanowires with $p+ip$-wave superconducting pairing have been investigated\cite{Potter,Niu,Rieder,Kells,Asahi,Zhou,Manmana,DeGottardi,Wimmer,Gangadharaiah}.
In the quasi-one-dimensional $p+ip$-wave pairing system, one pair of Majorana fermions appears if an odd number of subbands are filled.
This is because the class of the system is D, and is characterized by the $\mathbb{Z}_2$ index.
Recently, it was pointed out\cite{Tewari} that the realistic spin-orbit nanowire system approximately possesses chiral symmetry and the system belongs to the BDI class, where the number of MZES can have arbitrary integer values.

To realize the Kitaev model, there is a proposal to use a nanowire which has a strong spin-orbit interaction (SOI) with an ordinary $s$-wave superconductor and the Zeeman field. Because of the strong SOI, the system becomes a helical state, in which the electron spin and the momentum correspond one to one. The magnetic field perpendicular to the spin-orbit effective magnetic field breaks the time-reversal symmetry and opens a gap at $k=0$. Then, the system effectively becomes spinless when the Fermi energy is within this energy gap. Furthermore, the $s$-wave superconductor proximity is induced in the nanowire generates $p$-wave pairings.
A most promising candidate for a strong SOI nanowire is that of InSb, which has a large SOI ($\Delta_{\text{SOI}}=0.3\text{meV}$), a large electron $g$ factor ($|g| \sim 50$), and a small effective mass ($m^*=0.015m_e$).
Recently, the signature of  zero-energy bound states was observed in the system using InSb or InAs nanowire\cite{Mourik,Das,Deng,Rokhinson}.
However, experimental results are still controversial because there are many possible reasons for the zero-bias anomaly, such as the Kondo effect, disorder, or tunneling via Andreev bound states, which do not originate in Majorana fermions. \cite{EJHLee1,EJHLee2,Liu2,Rainis}

\begin{figure}[t]
\centerline{\includegraphics[width=0.4\textwidth]{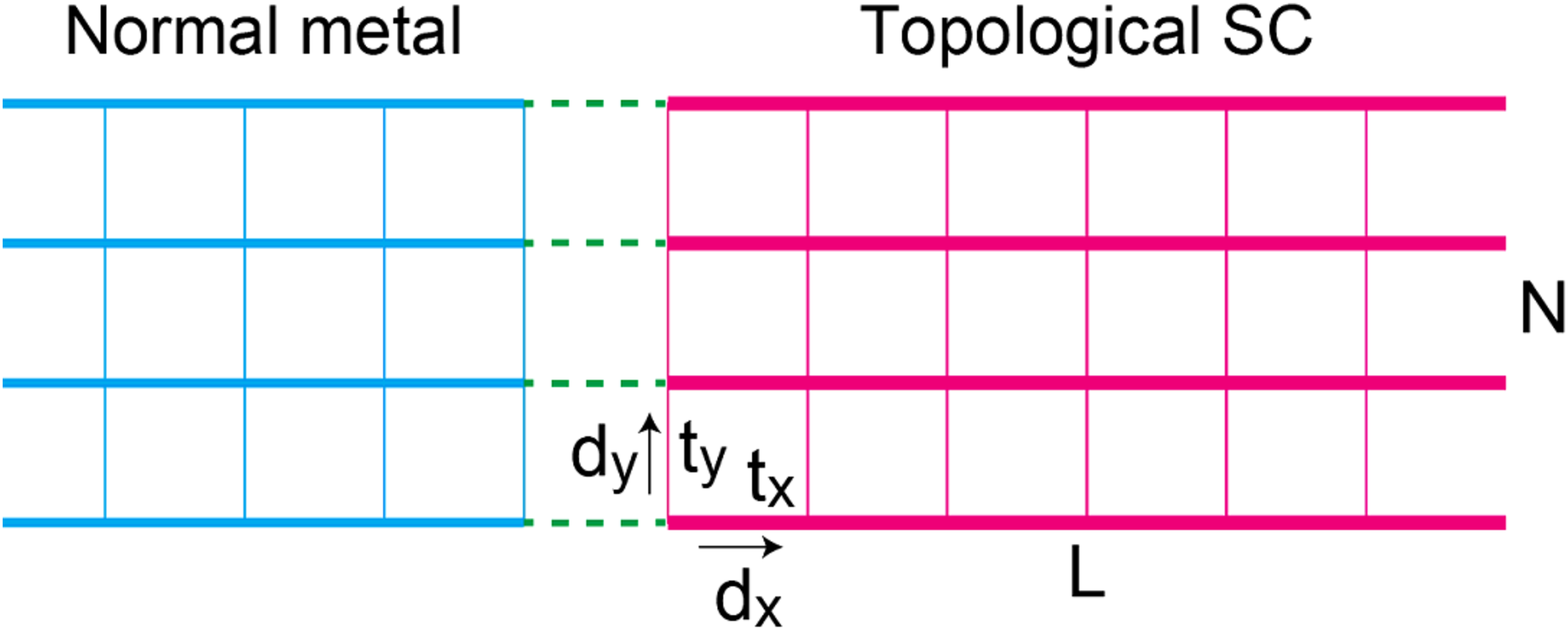}}
\caption{(Color online) Illustration of a ladder topological superconductor. It is
decomposed into the superconducting region (purple) and the lead (blue).
The purple region is the ladder topological superconductor.
Thick lines indicate the chains, while the thin lines indicate the interchain couplings. $L$ is
the number of sites along the chains, while $N$ is the number of the Kitaev chains. The blue region is the normal lead, which is attached when we calculate the differential conductance of the NS junction.}
\label{ladder}
\end{figure}

In this paper, motivated by the InSb nanowire system, we investigate a system where one-dimensional topological superconductors are placed in parallel, where the superconducting pairing phase can be changed arbitrarily.
The system would well be simulated by the ladder of topological superconductors shown in Fig. \ref{ladder}.
This model belongs to the class BDI or D, depending on the phase of the interchain pairing.
We first analyze the system where the interchain superconducting coupling is absent.
The system belongs to the class BDI, where the topological number is characterized by the $\mathbb{Z}$ index.
We find that the number of the MZES changes from $N$ to $0$ by changing the magnitude of the interlayer hopping.
We determine the topological phase diagram by calculating the $\mathbb{Z}$ index, which is a topological number of the system.
Next we introduce the interchain superconducting pairings.
When the relative phase between the inter- and intrachain superconducting pairings is $0$ or $\pi$, the system belongs to the class BDI and is characterized by the $\mathbb{Z}$ index.
Otherwise, the system becomes class D and is characterized by the $\mathbb{Z}_2$ index:
The number of the MZES changes alternately between $0$ and $1$ due to the level repulsion by the interchain superconducting pairings.

The rest of the paper is organized as follows.
In Sec .II we introduce the model and discuss the relationship between our model and the Kitaev model.
We also discuss the symmetry and the topological classification of the Hamiltonian.
In Secs .III and IV, we study the model without and with interchain superconducting pairing.
We calculate the $\mathbb{Z}$ and $\mathbb{Z}_2$ indices and determine the phase diagram.
We also discuss the energy spectrum.
In Sec. V we derive the lowenergy effective Hamiltonian written in terms of Majorana operators and discuss its low energy behavior.
In Sec. VI, examining the transport property of the model,
we study how the conductance depends on the phase difference between intra- and interchain superconducting pairings.

\section{Kitaev Ladder Model}

Majorana fermions emerge at the edges of an InSb nanowire in the presence of the superconducting order and the external magnetic field perpendicular to it.
It is well described by the Kitaev chain\cite{Kitaev}, where a one-dimensional $p$-wave topological superconductivity is realized.
A natural question is what happens when we place several InSb nanowires in parallel on or under the $s$-wave superconductor.
This setup will be easily realized using recent nanofabrication technology.

To simulate this system, we propose to investigate the Hamiltonian describing a ladder of the Kitaev chains,
\begin{align}
H=& -\mu \sum_{i,j}c_{i,j}^{\dagger }c_{i,j}  \notag \\
& -\sum_{i,j}(t_{x}c_{i,j}^{\dagger }c_{i+1,j}+t_{y}c_{i,j}^{\dagger
}c_{i,j+1}+\text{H.c.})  \notag \\
& -\sum_{i,j}(d_{x}c_{i,j}^{\dagger }c_{i+1,j}^{\dagger
}+d_{y}c_{i,j}^{\dagger }c_{i,j+1}^{\dagger }+\text{H.c.}),
\label{BasicHamil}
\end{align}
where $i$ and $j$ are the lattice coordinates for the $x$ and $y$ axes.
The intrachain transfer integral $t_{x}$ and the intrachain superconducting
pairing amplitude $d_{x}$ are present in the Kitaev model,
while we have newly introduced the effective interchain transfer integral $t_{y}$ and the
superconducting pairing amplitude $d_{y}$.

Physically, $t_y$ and $d_y$ will be derived by the couplings between the InSb nanowires via the substrate superconductor by the second-order process.
The superconducting pairing phase may be controlled by a superconducting quantum interference devide (SQUID) configuration.
The controllable phase difference between $d_x$ and $d_y$ is the new feature compared with the $p+ip$-wave pairing system.
We shall soon see that this superconducting pairing phase difference causes the topological class change between BDI and D.

There is an important comment on the coefficients $d_x$ and $d_y$.
By making a phase transformation of $c_{i,j}$, we can change the overall phase of the superconducting pairings.
Namely, the phase difference between $d_x$ and $d_y$ is physical, but the absolute phases of $d_x$ and $d_y$ are meaningless.
Without loss of generality we may assume that $d_x$ is real in the effective Hamiltonian, while $d_y$ is complex in general.
We define $\theta$ as the phase of $d_y$, i.e., $d_y=|d_y|e^{i\theta}$.

The energy spectrum is symmetric around zero energy due to the particle-hole symmetry induced by the superconductivity.
There are $N$ legs in the ladder (Fig. \ref{ladder}).
We refer to $L$ as the number of sites along  the ladder.
We note that the system is reduced to the Kitaev chain when $N=1$.
If the phase difference between $d_x$ and $d_y$ is $\pi/2$ and $L\approx N$, it is an anisotropic $p+ip$ superconductor system.
We analyze mainly the small-$N$ region.

We clarify the topological class\cite{Ryu} of the Hamiltonian (\ref{BasicHamil}).
In the momentum representation, the Bogoliubov--de Gennes Hamiltonian is given by
\begin{equation}
H(k)=
\begin{pmatrix}
\xi & \Delta \\
\Delta^\dagger & -\xi
\end{pmatrix},
\end{equation}
where $\xi$ and $\Delta$ are $N \times N$ tridiagonal matrices with the matrix elements
\begin{align}
\xi_{i,j} &= \left( -\mu - 2 t_x \cos{k} \right) \delta_{i,j} - t_y \delta_{i,j+1} - t_y \delta_{i+1,j}, \\
\Delta_{i,j} &= 2 id_x \sin{k} \delta_{i,j} + d_y \delta_{i,j+1} - d_y \delta_{i+1,j},\label{DeltaIJ}
\end{align}
where $k$ is the crystal momentum along the chain direction, i.e., the $x$ axis.

The explicit representations of the time reversal ($\Theta$), the particle-hole reversal ($\Xi$), and the chiral ($\Pi$) operator are defined by
\begin{equation}
\Theta=K, \quad \Xi=\tau_x K, \quad \Pi=\tau_x,
\end{equation}
where $K$ represents the complex conjugate operator.
We can show the Hamiltonian satisfies the relation
\begin{align}
&\Xi H(k) \Xi^{-1} = - H(-k), \\
&\Pi H(k) \Pi^{-1} = - H(k).
\end{align}
When Im$d_y=0$, the Hamiltonian satisfies the relation
\begin{equation}
\Theta H(k) \Theta^{-1} = H(-k).
\label{TRS}
\end{equation}
It is easily seen that $\Theta^2= \Pi^2= \Xi^2=1$.
Then, the Hamiltonian belongs to the class BDI, which is characterized by the $\mathbb{Z}$ index.
On the other hand, when Im$d_y\not=0$, the time-reversal symmetry is broken,
and the class changes to the class D where only $\Xi$ symmetry is present, being characterized by the $\mathbb{Z}_2$ index.

It is convenient to use the Majorana representation, which is constructed with use of the unitary transformation,
\begin{equation}
U=\frac{1}{\sqrt{2}}
\begin{pmatrix}
\mathbb{I}  & i\mathbb{I}  \\
\mathbb{I}  & -i\mathbb{I}
\end{pmatrix},
\end{equation}
where $\mathbb{I}$ is the unit matrix.
It transforms Dirac fermions into Majorana fermions,
\begin{equation}
c_{i,j}^\dagger=\frac{\gamma_{i,j}^A + i \gamma_{i,j}^B}{2}, \quad
c_{i,j}=\frac{\gamma_{i,j}^A - i \gamma_{i,j}^B}{2}, \quad
\label{GammaC}
\end{equation}
obeying $\left\{ \gamma_{i,j}^\alpha, \gamma_{i',j'}^{\alpha'} \right\} = \delta_{i,i'} \delta_{j,j'} \delta_{\alpha,\alpha'}$.
The unitary transformed Hamiltonian is given by
\begin{equation}
U^\dagger H U = \frac{1}{2}
\begin{pmatrix}
\Delta + \Delta^\dagger & 2 i \xi - i \left( \Delta - \Delta^\dagger \right) \\
-2 i \xi - i \left( \Delta - \Delta^\dagger \right) & - \left( \Delta + \Delta^\dagger \right)
\end{pmatrix}.
\end{equation}
Our analysis is carried on based on this Majorana representation of the Hamiltonian.

\section{topological phase diagram without interchain superconducting pairings}

We first investigate the case of $d_y=0$ for simplicity.
The transformed Hamiltonian is written as
\begin{equation}
U^\dagger H U = i
\begin{pmatrix}
0 & v \\
-v^T & 0
\end{pmatrix},
\label{HamilU}
\end{equation}
which has only off-diagonal elements, with
\begin{align}
v&= \xi - \frac{1}{2} \left( \Delta - \Delta^\dagger \right) \nonumber\\
&= \left( -\mu - 2 t_x \cos{k} - 2 i d_x \sin{k} \right) \mathbb{I} - t_y \mathbb{M}, \label{MatV}
\end{align}
where $\mathbb{M}$ represents a tridiagonal matrix which only has elements $1$ at $(n,n+1)$ and $(n+1,n)$.
It is straightforward to diagonalize the matrix $v$.
As we see at the end of this section, the eigenvalues are
\begin{equation}
z_n= -\mu - 2 t_y \cos{\frac{n\pi}{N+1}} - 2 t_x \cos{k} - 2 i d_x \sin{k},
\label{EigenZn}
\end{equation}
while the eigenfunctions are
\begin{equation}
\Psi_n =\sqrt{\frac{2}{N+1}} \sin \frac{n m \pi}{N+1} \quad (n=1,2,\cdots,N),
\label{EigenFn}
\end{equation}
where $m$ represents the position ($m=1,2,\cdots,N$).

\begin{figure}[t]
\centerline{\includegraphics[width=0.4\textwidth]{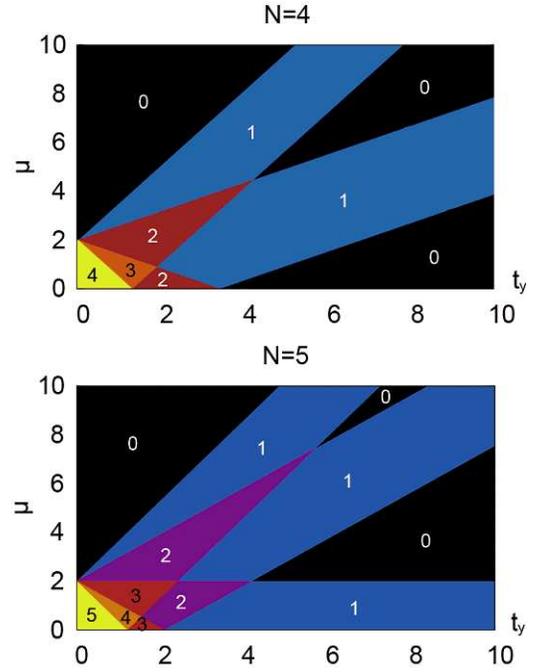}}
\caption{(Color online) Phase diagrams of the four-leg and five-leg ladders.
The energy unit is $t_x$, and we have taken $d_x=0.5$.
The numbers in the figure represent the $\mathbb{Z}$ index. The $\mathbb{Z}_2$ index is the parity of the number.}
\label{PD}
\end{figure}

The $\mathbb{Z}$ index $W$ is defined by the chiral index
\begin{align}
W &\equiv \int \frac{dk}{2 \pi i} \partial_k \ln \det v \left( k \right) \notag \\
&=\mathrm{tr} \int \frac{dk}{2 \pi i} \partial_k \ln{v\left( k \right)}
=\sum_n \int \frac{dk}{2 \pi i} \partial_k \ln{z_n\left( k \right)},
\label{index}
\end{align}
where $z_n$ are the eigenvalues of $v$.
It describes the sum of the winding numbers of $v(k)$ around the origin of the complex plane.
The $\mathbb{Z}$ index is calculable by substituting (\ref{EigenZn}) into (\ref{index}).

We investigate the topological phase diagram in the $(t_y,\mu)$ plane.
The phase boundaries are determined by $\text{det}v=0$.
It follows from (\ref{EigenZn}) that they are given by
\begin{equation}
-\mu-2 t_y \cos{\frac{n \pi}{N+1}}=\pm 2 t_x.
\label{EqA}
\end{equation}
The topological phase diagram is illustrated in Fig. \ref{PD}.
We see the following characteristic features:
The $\mathbb{Z}$ index is $W=N$ at $(t_y,\mu) =(0,0)$.
Along the $\mu$ axis with $t_y=0$, the topological phase transition occurs from $W=N$ to $W=0$ at $\mu =\pm 2t_x$ irrespective of the number $N$ of the legs.
On the other hand, along the $t_y$ axis with $\mu=0$,
multiple topological phase transitions occur:
The band gap closes at $t_y=\pm \frac{2 t_x}{z_n}$, and
the $\mathbb{Z}$ index is $0$ for even $N$ and $1$ for odd $N$ for sufficiently large $t_y$.

\begin{figure}[t]
\centerline{\includegraphics[width=0.4\textwidth]{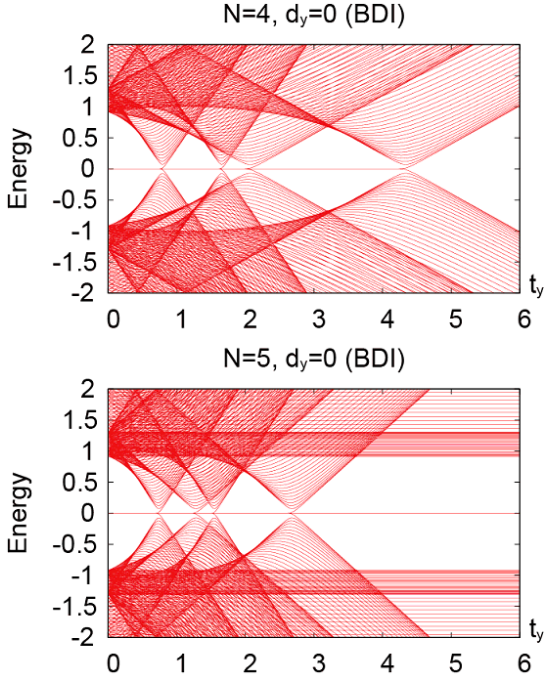}}
\caption{(Color online) Energy spectra of an $N=4,5$ ladder with $d_y=0$.
The energy unit is $t_x$, and we have taken $\mu=0.7$, $d_x=0.5$.
The number of zero-energy states decreases
by 2 at every phase transition as $t_y$ increases,
since each zero-energy state per chain doubly degenerates due to the two end points of the chain. In the strong coupling limit $t_y \rightarrow \infty$, there is no (one) MZES per each edge in case of $N=4 (5)$. We have set $L=64$.}
\label{EneZ}
\end{figure}

We show the energy spectrum of the ladder of chains as a function of $t_y$ for $N=4,5$ in
Fig. \ref{EneZ}.
We find that the $\mathbb{Z}$ index gives the number of MZES.
The wave functions of all the MZES are localized at the edges.

The above behaviors can be understood as follows.
The system is equivalent to $N$ independent Kitaev chains with the renormalized chemical potential:
\begin{align}
H&=\sum_{n=1}^N H_n, \\
H_n&=-\mu_n' \sum_i c_{i,n}^\dagger c_{i,n} \notag \\
& \quad - \sum_i \left( t_x c_{i,n}^\dagger c_{i+1,n} + d_x c_{i,n}^\dagger c_{i+1,n}^\dagger + \text{H.c.} \right), \\
\mu_n'& = \mu + 2 t_y \cos{\frac{n}{N+1}\pi}.
\end{align}
Each renormalized Kitaev chain has MZES when $|\mu_n'| < 2 |t_x|$.
The total number of the MZES is given by the sum of the renormalized Kitaev chains.

For the sake of completeness we show how to derive the eigenvalues (\ref{EigenZn}).
The eigenvalues $z_n$ can be obtained by solving the characteristic equation for $t_y \mathbb{M}$,
\begin{equation}
f_N(x)=\mathrm{det}\left( x\mathbb{I} - t_y \mathbb{M} \right)=0.
\end{equation}
To solve this we write down the recurrence relation of the characteristic equation,
\begin{equation}
f_{N+2}(x)=x f_{N+1}(x) - t_y^2 f_N(x),
\label{recur}
\end{equation}
from which it follows that
\begin{equation}
f_N(x)=t_y^N U_N \left( \frac{x}{2t_y} \right)=\sum_{n=0}^{\left[ N/2 \right]} \left( -1 \right)^n \binom {N-n}{n} t_y^{2n} x^{N-2n}
\end{equation}
with Chebyshev polynomials of the second kind.
We now solve the eigenequation as
\begin{equation}
x_n = 2 t_y \cos{\frac{n \pi}{N+1}} \quad (n=1,2,\cdots,N),
\end{equation}
from which we find the eigenvalues (\ref{EigenZn}) and the eigenfunctions (\ref{EigenFn}) because $z_n$ is constant minus $x_n$.

\section{topological phase diagram with interchain superconducting pairings}

\begin{figure}[t]
\centerline{\includegraphics[width=0.4\textwidth]{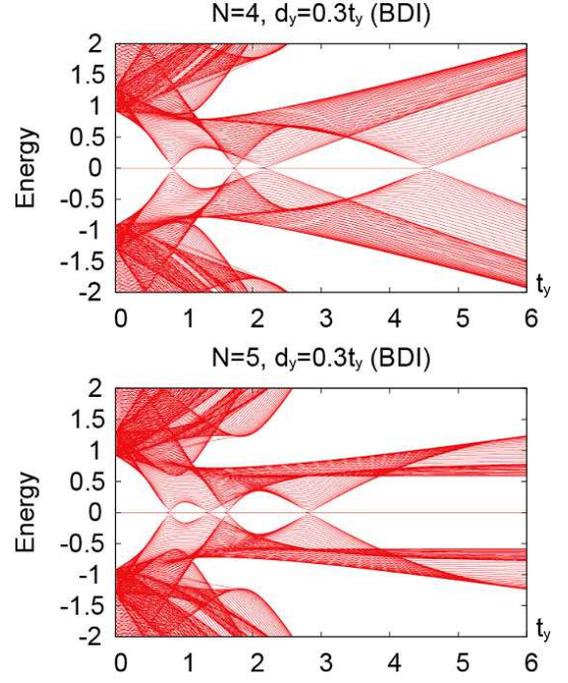}}
\caption{(Color online) Energy spectra of $N=4,5$ ladder with $d_y=0.3t_y$.
The energy unit is $t_x$, and we have taken $\mu=0.7$, $d_x=0.5$.
Note that both $d_x$ and $d_y$ are real. The number of zero-energy states decreases by two at every topological phase transition with the gap closing of the bulk states,
because each zero-energy state per chain doubly degenerates due to the two endpoints of the chain.
In the strong coupling limit $t_y \rightarrow \infty$, there is no (one) MZES per each edge in case of $N=4 (5)$. We have set $L=64$.}
\label{EneBDI}
\end{figure}

We next investigate the case $d_y\not=0$.
We first assume that it is real as well as $d_x$.
Then, the system belongs to the class BDI,
and the system is characterized by the $\mathbb{Z}$ index.
We show the energy spectrum in Fig. \ref{EneBDI}.

We proceed to investigate the effects of complex interchain superconducting pairings (Im$d_y\not=0$).
The Hamiltonian no longer satisfies the time-reversal symmetry (\ref{TRS}).
The topological class of the Hamiltonian changes to the class D, and the topological number is the $\mathbb{Z}_2$ index.
The system allows only the two cases, whether or not a pair of Majorana fermions exist.

\begin{figure}[t]
\centerline{\includegraphics[width=0.4\textwidth]{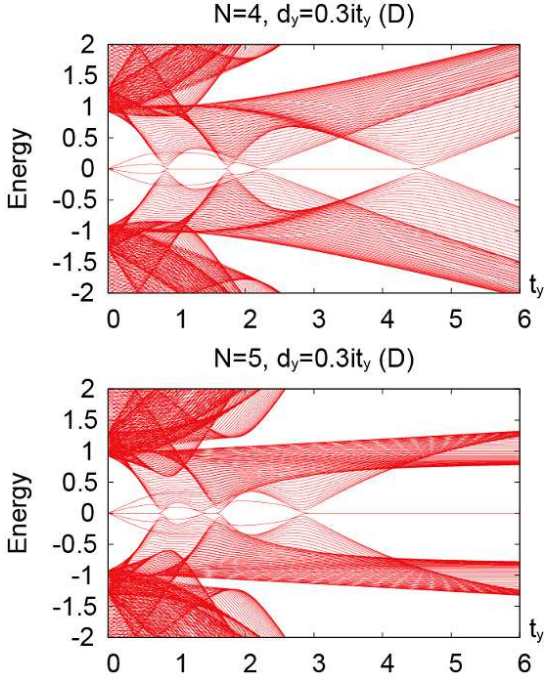}}
\caption{(Color online) Energy spectra of $N=4,5$ ladder with $|d_y|<t_y$.
The energy unit is $t_x$, and we have taken $\mu=0.7$, $d_x=0.5$, and $d_y=0.3 i t_y$.
Each subgap state doubly degenerates.
With Im$d_y\not=0$, MZES disappear in the parameter region where there are even MZES for $d_y=0$, while two MZES remain in the parameter region where there are odd MZES for $d_y=0$. In the strong coupling limit $t_y \rightarrow \infty$, there is no (one) MZES per each edge in case of $N=4 (5)$. We have set $L=64$.}
\label{EneD}
\end{figure}

We show the energy spectrum as a function of $t_y$ in the presence of $d_y$ (Fig. \ref{EneD}).
The behavior of the MZES drastically changes from the system with $d_y=0$ to the one with Im$d_y\not=0$.
When we change the phase of $d_y$ and the class changes from BDI to D, the number of MZES reduces from even (odd) number to zero (one).
This is due to the interference of MZES pairs due to the interchain superconducting pairings.
Then, only one MZES pair can exist, which implies that the topological number is the $\mathbb{Z}_2$ index.
The number of MZES pairs is given by $\text{mod}_2 N$.

The $\mathbb{Z}_2$ index is obtained by the sign of the Pfaffian at the time-reversal-invariant momenta $k=0$ and $\pi$.
At these points, since $d_x$ disappears from $\Delta$ defined by (\ref{DeltaIJ}),
it is convenient to make a phase transformation
$c \mapsto e^{i \theta / 2} c,$
so that $d_y$ becomes real.
Then, the Majorana represented Hamiltonian at these points is given by (\ref{HamilU}), i.e.,
\begin{equation}
U^\dagger H U = i
\begin{pmatrix}
0 & v_\pm \\
-v_\pm^T & 0
\end{pmatrix}
\end{equation}
but now with
\begin{align}
v_\pm&=\left( \xi_\pm - \Delta \right) \notag \\
&= \left( -\mu \pm 2 t_x \right) \delta_{i,j}
- \left( t_y + |d_y| \right) \delta_{i,j+1} - \left( t_y - |d_y| \right) \delta_{i+1,j},
\end{align}
where $\xi_-=\xi(0)=-\mu-2t_x$ and $\xi_+=\xi(\pi)=-\mu+2t_x$.

\begin{figure}[t]
\centerline{\includegraphics[width=0.4\textwidth]{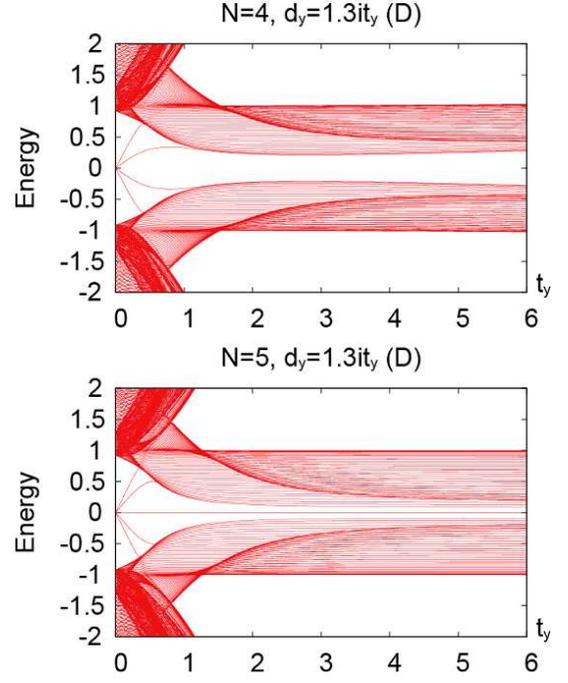}}
\caption{(Color online) Energy spectra of $N=4,5$ ladder with $|d_y|> t_y$.
The energy unit is $t_x$, and we have taken $\mu=0.7$, $d_x=0.5$, and $d_y=0.3 i t_y$.
Each sub-gap state doubly degenerates. When $N=4(5)$, the system is always trivial (topological). We have set $L=64$.}
\label{EneL}
\end{figure}

The Pfaffian is simply given by
\begin{equation}
\text{Pf}
\begin{pmatrix}
0 & v_\alpha \\
-v_\alpha^T & 0
\end{pmatrix}=\pm \text{det}v_\alpha, \label{Pfaffian}
\end{equation}
with $\alpha=\pm$, since there are no diagonal elements.
Then $\mathbb{Z}_2$-invariant is the sign of the product of the Pfaffians in Eq. (\ref{Pfaffian}), i.e., $\text{det}v_+ \text{det}v_-$.
We note that when $d_y$ is real, the $\mathbb{Z}_2$-invariant is the parity of the $\mathbb{Z}$-invariant.
The topological phase boundaries are determined by $\text{det}v=0$ as before.

We obtain the recursion relation of the characteristic polynomial $f_N=\text{det} v$ as
\begin{equation}
f_{N+2} = \left( -\mu \pm 2t_x \right) f_{N+1} - \left( t_y^2 - |d_y|^2 \right) f_{N}.
\end{equation}
This relation is obtained by replacing $t_y^2$ to $t_y^2-|d_y|^2$ in (\ref{recur}).
Then, the condition $f_{N}=0$ yields
\begin{equation}
\mu-2 \sqrt{t_y^2-|d_y|^2} \cos{\frac{n \pi}{N+1}} =\pm 2 t_x,
\end{equation}
which is nothing but (\ref{EqA}) with the replacement of $t$ by $t_y^{\prime}=\sqrt{t_y^2-|d_y|^2}$.
The phase diagram depends on only $|d_y|$.

We obtain the phase diagram as follows.
When $|t_y|>|d_y|$, it is simply given by Fig. \ref{PD} with the understanding that the horizontal axis is $t_y^{\prime}$.
Recall that the energy spectrum as a function of $t_y$ is given by Fig. \ref{EneD}.
On the other hand, when $|d_y| > t_y$, $t_y'$ becomes pure imaginary,
and the phase diagram becomes very simple:
When $N$ is odd, the gap closing condition is given only by $\mu = \pm 2 t_x$.
When $N$ is even, on the other hand, the system is trivial in the whole range.
The energy spectrum as a function of $t_y$ is now given by Fig. \ref{EneL}.

\begin{figure}[t]
\centerline{\includegraphics[width=0.4\textwidth]{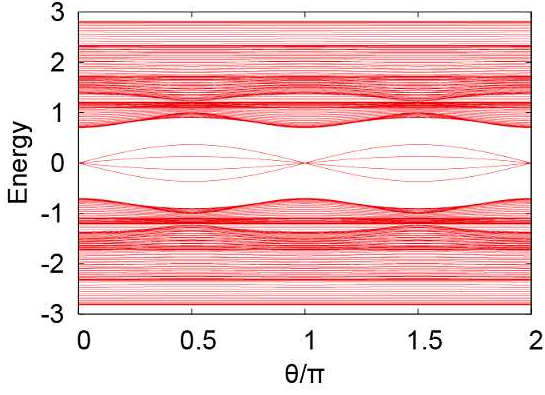}}
\caption{(Color online) Energy spectrum as a function of the superconducting pairing phase $\theta$.
The energy unit is $t_x$, and we have taken
$\mu=0$, $d_x=0.5$, $t_y=0.5$, $d_y=0.25$, and $L=64$.
There are four MZES in the case of $N=4$, which are proportional to $\sin \theta$.}
\label{Enephase}
\end{figure}

\section{Energy spectrum as a function of the superconducting phase difference}

It is interesting to investigate how the energy spectrum changes during the topological class change between the BDI class and the D class.
Therefore, we show the energy spectrum as a function of $\theta$ defined by $d_y =|d_y|e^{i\theta}$ in Fig. \ref{Enephase}.
One can clearly see the MZES is clearly seen in the bulk band gap.
Here, we derive the $\sin \theta$ dependence of the subgap levels.
In the Kitaev model, if we define the Majorana operators as in Eq. (\ref{GammaC}), the zero-energy Majorana fermion on the left (right) edge contains only $\gamma^A$ ($\gamma^B$) as in Kitaev's original paper\cite{Kitaev}.
Therefore, at low energy, the $d_y$ term in our Hamiltonian is reduced to
\begin{align}
& |d_y| e^{i \theta} c_{i,j}^\dagger c_{i,j+1}^\dagger + \text{H.c.} \notag \\
&\qquad = -\frac{i |d_y|}{2} \cos \theta \left( \gamma^A_{i,j} \gamma^B_{i,j+1} + \gamma^B_{i,j} \gamma^A_{i,j+1} \right) \notag \\
&\qquad \quad -\frac{i |d_y|}{2} \sin \theta \left( \gamma^A_{i,j} \gamma^A_{i,j+1} - \gamma^B_{i,j} \gamma^B_{i,j+1} \right) \\
&\qquad \approx
\begin{cases}
- \frac{i |d_y|}{2} \sin \theta \gamma^A_{i,j} \gamma^A_{i,j+1} \quad (i \in \text{left half}) \\
+ \frac{i |d_y|}{2} \sin \theta \gamma^B_{i,j} \gamma^B_{i,j+1} \quad (i \in \text{right half})
\end{cases}
.
\end{align}
Namely, the Majorana fermions at chain $j$ and $j+1$ are coupled by the imaginary part of the interchain pairing, and this causes the $\sin \theta$ dependence of the energy:
\begin{equation}
E \propto |d_y| \sin \theta.
\end{equation}
In other words, breaking of the time-reversal symmetry by the imaginary part of the interchain coupling corresponds to the creation of the coupling between Majorana fermions.
By numerically diagonalizing the Hamiltonian (\ref{BasicHamil}), we obtain the energy spectrum as a function of $\theta$ for $d_x/t_x=0.5$ as shown in Fig. \ref{Enephase}.
The subgap levels are in good agreement with the sinusoidal behavior.
We have thus discussed the $\sin \theta$ dependence of the subgap levels.
An effective Hamiltonian on the subgap states will be derived in a manner similar to that in Ref. \cite{Kells}.
We note that the topological class change without gap-closing occurs when we change $\theta$\cite{Ezawa}.

\section{transport property}

It is well known\cite{Law,Fidkowski} that the local Andreev reflection rate at zero bias is $1$ in the presence of a Majorana fermion, while the zero-energy Majorana bound state gives the differential conductance $2e^2/h$.
We have calculated the differential conductance of the NS junction (Fig. \ref{ladder}), following Refs.\cite{LeeFisher,FisherLee,He,Liu,Ii,Lewenkopf}.
By employing the recursive Green's function method, we obtain the surface Green's function\cite{Sancho,Matsumoto,Umerski}
of the semi-infinite Kitaev ladder numerically.
In the Matsumoto-Shiba formalism\cite{Matsumoto}, we express the semi-infinite wire by the delta-function potential with infinite strength.
Then, the Green's function of the semi-infinite wire $G_{i,i'}$ is given by
\begin{equation}
G_{i,i'} = G^0_{i,i'} - G^0_{i,0} \left( G^0_{0,0} \right)^{-1} G^0_{0,i'},
\end{equation}
where $G^0_{i,i'}$ is the bulk Green's function.
The bulk Green's function is given analytically in $k$-space.
Then the real-space representation is obtained by performing Fourier transformation numerically:
\begin{equation}
G^0_{i,i'} = \frac{1}{2\pi} \int dk e^{i k \left( i-i' \right)} G^0 (k).
\end{equation}
On the other hand, the surface Green's function of the semi-infinite normal lead can be given analytically:
\begin{align}
 g_{m,n} (E)=
\begin{cases}
A(k)\left( E_J-i \sqrt{(2t)^2-E_J^2} \right) & \left( |E_J|<2t \right) \\
A(k)\left( E_J+\sqrt{E_J^2-(2t)^2} \right) & \left( E_J<-2t \right) \\
A(k) \left( E_J-\sqrt{E_J^2-(2t)^2} \right) & \left( E_J>2t \right)
\end{cases},
\end{align}
where $m$ and $n$ are the chain labels;
$A(k)=\frac{2}{N+1} \frac{1}{2t^2} \sum_k \sin (mka) \sin (nka) $; and
$E_J=E+\mu-2t_y \cos ka$, where $t(>0)$ and $\mu$ are the hopping and the chemical potentials of the leads.

Then, we construct the Green's function of the whole system by the recursion relations.
Expressing the Green's function of the left(right) semi-infinite wire as $G_L(G_R)$ and the Green's function of the whole system as $G$, we obtain
\begin{align}
&G_{L,i,i}^{-1} = g_i^{-1} - H_{i,i-1} G_{L,i-1,i-1} H_{i-1,i}, \\
&G_{R,i,i}^{-1} = g_i^{-1} - H_{i,i+1} G_{R,i+1,i+1} H_{i+1,i}, \\
&G_{i,i}^{-1} = g_i^{-1} - H_{i,i-1} G_{L,i-1,i-1} H_{i-1,i} \notag \\
& \qquad \qquad \qquad \qquad - H_{i,i+1} G_{R,i+1,i+1} H_{i+1,i}, \\
&G_{i,i+1} = G_{i,i} H_{i,i+1} G_{R,i+1,i+1}, \\
&G_{i+1,i} = G_{i+1,i+1} H_{i+1,i} G_{L,i,i}, \\
&g_i^{-1} = E - H_i,
\end{align}
where $E$ is the energy, $H_i$ is the on-site Hamiltonian, and $H_{i,i'}$ is the hopping between sites $i,i'$.
On the other hand, we obtain the retarded Green's function $G^R$ by replacing $E$ with $E + i \varepsilon$, where $\varepsilon$ is an infinitesimal positive number.

Next, we calculate the differential conductance by the Lee-Fisher formula\cite{LeeFisher,Ii}:
\begin{align}
G=\frac{2e^2}{h} \text{Tr} \left[ \right. P_\text{e} \left( \right. &G''_{i,i+1} G''_{i,i+1} + G''_{i+1,i} G''_{i+1,i} \notag \\
& - G''_{i,i} G''_{i+1,i+1} - G''_{i+1,i+1} G''_{i,i} \left. \right) \left. \right],
\end{align}
where $G''_{i,i'} = \text{Im} G^R_{i,i'}$, $P_\text{e}$ is the projection operator onto the particle subspace.
We choose an arbitrary $i$ in the normal region due to current conservation.

\begin{figure}[t]
\centerline{\includegraphics[width=0.4\textwidth]{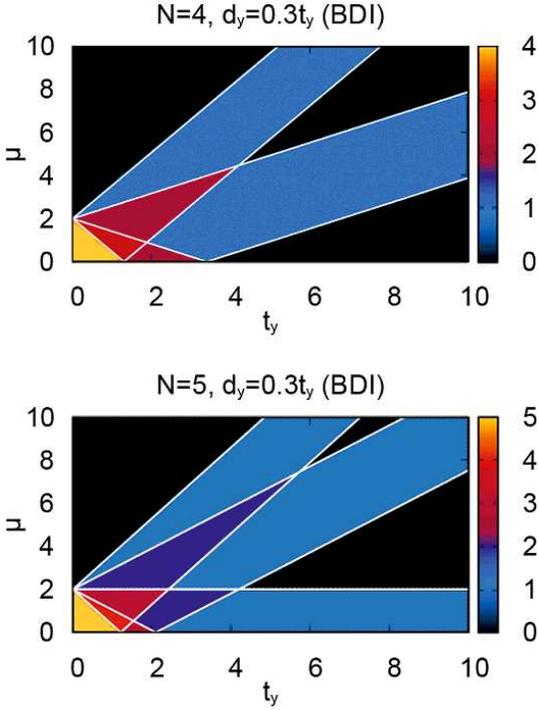}}
\caption{(Color online) Differential conductance of the NS junction in units of $2e^2/h$ with real $d_y$ when $N=4,5$. The system belongs to the class BDI.
The energy unit is $t_x$, and we have taken
$d_x=0.5$ and $d_y=0.3t_y$. White lines are topological phase boundaries in Fig. \ref{PD}.}
\label{CndBDI}
\end{figure}

\begin{figure}[t]
\centerline{\includegraphics[width=0.4\textwidth]{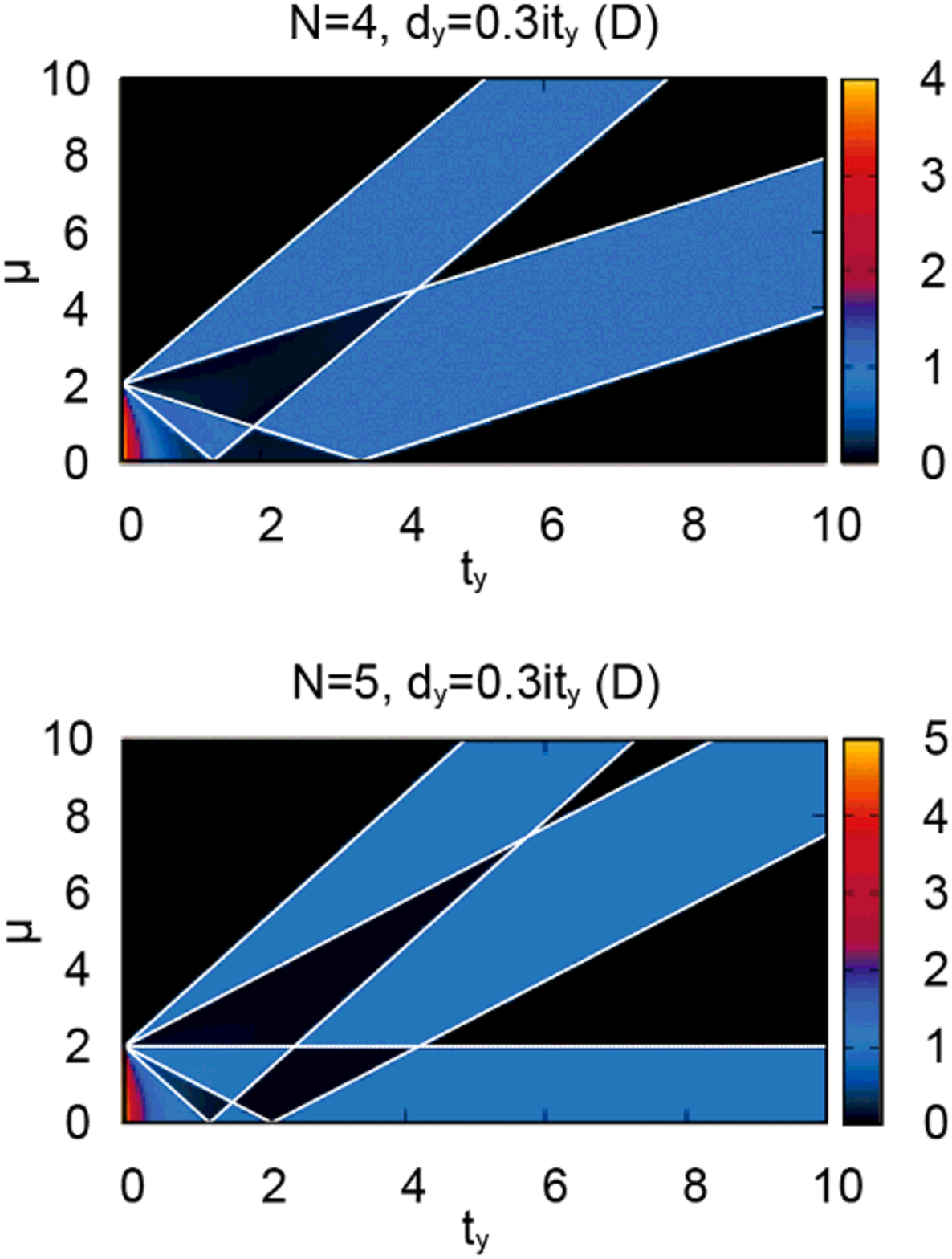}}
\caption{(Color online) Differential conductance of the NS junction in units of $2e^2/h$ with imaginary $d_y$ when $N=4,5$. The system belongs to the class D.
The energy unit is $t_x$, and we have taken
$d_x=0.5$ and $d_y=0.3it_y$.
White lines are topological phase boundaries in Fig. \ref{PD}.}
\label{CndD}
\end{figure}

\begin{figure}[t]
\centerline{\includegraphics[width=0.4\textwidth]{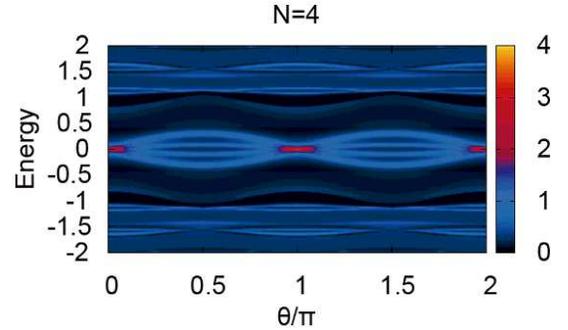}}
\caption{(Color online) Differential conductance of the NS junction in units of $2e^2/h$ as a function of the superconducting pairing phase $\theta$ and the energy.
The energy unit is $t_x$, and we have taken
$\mu=0$, $d_x=0.5$, $t_y=0.5$, and $d_y=0.25$.
There are four MZES in the case of $N=4$.}
\label{Cndphase}
\end{figure}

The results for the differential conductance $G$ at zero energy are illustrated in Figs.\ref{CndBDI} and \ref{CndD}.
We have investigated the cases with real $d_y$ (class BDI) and with imaginary $d_y$ (class D).
In both of the classes BDI and D, the figures of the conductance strongly resemble the phase diagrams (Fig. \ref{PD}).
The bright regions where the differential conductance is quantized correspond to the topological regions.
This means that the local Andreev reflection rate is 1 with each MZES.
It agrees with the previous study\cite{Romito,Law,Flensberg} that the Majorana fermion induces the resonant Andreev reflection.

We have calculated the differential conductance as a function of the superconducting pairing phase $\theta$ and the energy of injected electrons, which we show in Fig. \ref{Cndphase}.
The behavior reflects the behavior of the energy spectrum given in Fig. \ref{Enephase}.
Moreover, the differential conductance per mode is quantized to $2e^2/h$ when $\theta=0,\pi$.
This is because of the resonant Andreev reflection induced by Majorana fermions.
Furthermore, the peak decays rapidly when the phase is switched on and there are no Majorana fermions.

We have also calculated the crossed Andreev reflection rate in the NSN junction, where the two Majorana fermions couple.
The coupling strength between Majorana fermions at two edges is on the order of $\text{exp}(-L/\xi)$, where $\xi$ is the superconducting coherence length.
When $\xi \ll L$, since there are no overlaps between the MZES localized at the right and left edges, the crossed Andreev reflection rate is quite small.
On the other hand, when $\xi >L$, there are some overlaps between the MZES localized at the right and left edges, and the crossed Andreev reflection rate remains finite.

\section{Conclusion and Discussion}

Motivated by the InSb nanowire superconductor system, we have investigated Kitaev-ladder topological superconductors based on the effective Hamiltonian (\ref{BasicHamil}).
Especially, we have revealed the topological phase transition by changing
the superconducting pairing phase difference between the $x$ and $y$ directions (i.e., the phase of $d_y/d_x$).
This is the problem not addressed in the previous studies\cite{Potter,Niu,Rieder,Kells,Asahi,Zhou,Manmana,DeGottardi,Wimmer,Gangadharaiah}
on the $p+ip$-wave pairing system.
We have found that the pairing phase plays a crucial role to determine the topological class.
If the phase is $0$ or $\pi$, the system belongs to the class BDI with $n$ MZES ($n=0,1,2,...$);
otherwise it  belongs to the class D with $n$ MZES ($n=0,1$).
The phase gradient of the bulk superconductor may be controlled by forming the SQUID configuration.
The proximity-induced pairing in the wires also gets a phase gradient, yielding an imaginary part to $d_y$,
and the class changes to D.
This topological class change can be observed by differential conductance measurement.
We note that our system is mapped to the Kitaev-spin-ladder system\cite{DeGottardi2,Pedrocchi} in the case of $d_y=0$.
It may be possible that our model is realized by using ultra-cold atomic systems\cite{Zhu}.

{\it Note added in proof.} We became aware of the nice papers on a similar system\cite{Wu1,Wu2,Beenakker2}

\section*{Acknowledgements}

We are very grateful to Y. Tanaka for many
helpful discussions on the subject.
This work was supported in part by Grants-in-Aid for Scientific Research
from the Ministry of Education, Science, Sports and Culture No. 24224009 and No. 25400317.

\end{document}